\documentclass[11pt,aps,pra,onecolumn,superscriptaddress,amssymb,amsfonts,showpacs,tightenlines]{revtex4}

\usepackage[dvips]{graphics,graphicx,epsfig}
\newcommand{\cmplxs}{{\mathbb C}}
\newcommand{\rlxs}{{\mathbb R}}

\begin{document}
\date{\today}

\title{Contractive Schr\"odinger cat states for a free mass}

\author{Lorenza Viola}
\email[Electronic address: ]{lviola@lanl.gov}  
\affiliation{Los Alamos National Laboratory, Los Alamos, NM 87545, USA}
\author{Roberto Onofrio}
\email[Electronic address: ]{onofrio@pd.infn.it} 
%\affiliation{Los Alamos National Laboratory, Los Alamos, NM 87545, USA}
\affiliation{Dipartimento di Fisica ``G. Galilei", 
Universit\`a di Padova, via Marzolo 8, 35131 Padova, Italy}
\affiliation{Istituto Nazionale di Fisica della Materia, Unit\`a di Roma 
``La Sapienza", and Center for Statistical Mechanics and Complexity, 
Piazzale A. Moro 2, Roma 00185, Italy}
\affiliation{Los Alamos National Laboratory, Los Alamos, NM 87545, USA}

\begin{abstract} 
Contractive states for a free quantum particle were introduced by 
Yuen (Yuen H P 1983 Phys. Rev. Lett. {\bf 51}, 719) in an attempt to evade 
the standard quantum limit for repeated position measurements. 
We show how appropriate families of two- and three component  
``Schr\"odinger cat states'' are able to support non-trivial 
correlations between the position and momentum observables leading to 
contractive behavior. 
The existence of contractive Schr\"odinger cat states is suggestive of 
potential novel roles of non-classical states for precision measurement 
schemes. 
\end{abstract}

%%%%%%%%%%%%%%%%%%%%%%%%%%%%%%%%% PACS NUMBERS
%\footnotetext{Keywords:}
% 04.80.-y Experimental studies of gravity
% 03.65.Ta Foundations of quantum mechanics; measurement theory
% 03.65.Yz Decoherence; open systems; quantum statistical methods
%%%%%%%%%%%%%%%%%%%%%%%%%%%%%%%%% PAPER CONTENT
\maketitle

\section{Introduction}

Shortly after the seminal paper by Schr\"odinger in 1935 \cite{SCHROEDINGER}, 
``Schr\"odinger cat'' became a pictorial way to refer to a prototypical 
family of genuinely non-classical states {\it i.e.}, quantum states without
a classical counterpart. While the introduction of such states found its 
original motivation in the analysis of the celebrated cat 
{\sl Gedankenexperiment} in quantum measurement theory 
\cite{SCHROEDINGER,WHEELER}, the quantum-mechanical weirdness of Schr\"odinger
cat states has since then been widely appreciated in various contexts. 
The latter range from quantum non-locality \cite{BELL}, to non-classical 
states of light and matter \cite{HAROCHE}, and to the emergence of 
classicality from the quantum world \cite{ZUREK,GIULINI}. 
In particular, cat states arising from the superposition of 
macroscopically distinguishable states of a quantum degree of freedom 
play a prominent role in the debate on macroscopic quantum mechanics 
\cite{LEGGETT}. More recently, the rapid development in the field of 
quantum information science \cite{CHUANG} has prompted the 
consideration of various classes of non-classical states 
from the point of view of the quantum resources that they may involve
\cite{PRESKILL,HUELGA,MEYER}. For instance, cat-like superpositions of 
bosonic coherent states have been shown to provide robust encodings
against amplitude damping \cite{COCHRANE}, to carry the potential for 
improving the sensitivity of weak-force detection \cite{MUNRO}, as well 
as to exhibit generalized entanglement properties \cite{BARNUM}.

While applications of non-classical states within quantum information 
processing are still emerging, their association with the field of 
high precision measurements has a long history. A leading motivation 
stems from the need for understanding the ultimate resolution limitations 
for repeated position measurements in experimental and observational 
gravitation \cite{THORNE}. 
Without a careful preparation of the initial state, quantum mechanics 
sets a limit based on the assignment of equal uncertainties to position 
and momentum, resulting in the so-called {\sl Standard Quantum Limit}
(SQL) \cite{BRAGINSKY67,INTERCAVES}. Attempts to beat the SQL have stimulated 
the development of a theory for quantum non-demolition measurements 
\cite{BRAGINSKY,CAVES}, and several experimental proposals aimed at 
improving the sensitivity of resonant detectors of gravitational radiation 
\cite{BOCKO}. Efforts have primarily focused on a harmonic oscillator
\cite{CAVES}, resulting in the possibility of overcoming the SQL using 
non-classical squeezed states. The case of a free test mass did not receive 
much attention until Yuen proposed a novel class of so-called 
{\sl contractive states} analogous to the bosonic two-photon 
coherent states \cite{YUEN} and capable, in principle, of narrowing
the position variance below the SQL bound.  
While Yuen's proposal initiated a controversial debate   
\cite{LYNCH,CAVES1,OZAWA,NI,BONDURANT,BRAGINSKY2001,OZAWA2002}, 
the demand for schemes able to beat the SQL for free masses has meanwhile 
increased due to the amazing improvements in the sensitivity of gravitational 
wave interferometric detectors \cite{INTERFER}. In principle, even a 
modest gain in the sensitivity obtainable by exploiting non-classical 
states would imply a relatively large increase in the fiducial volume 
for gravitational wave detection \cite{CAVESINT,ANSARI,KIMBLE}. 
Unfortunately, contractive states as proposed by Yuen are difficult 
to produce in practice; the only concrete proposal that has appeared so far 
in the literature was regarding their generation for atomic states \cite{STOREY}. 

In this paper, we revisit Schr\"odinger cat states from the 
motivating perspective of the SQL of a free mass, and show how in addition 
to their already known properties they are also capable of manifesting
 contractive features. The content is organized as follows. 
In Section II we summarize the basic notions about the 
contractivity property as introduced by Yuen. In Section III we 
investigate the behavior of the position variance for a paradigmatic 
class of Schr\"odinger cat states, evidencing its dependence upon various 
parameters, and then outlining possible directions for generalization of 
this result. Section IV is devoted to discussing some implications of the 
existence of contractive Schr\"odinger cat states, and preliminarily
assessing the prospects for demonstrating contractive cat behavior. 
Additional remarks are presented in the conclusions.

\section{ Contractivity and Yuen states }

Consider a quantum particle of mass $m$ freely evolving in one dimension.
If $\hat{x}, \hat{p}$ are the position and momentum observables, the 
equation of motion for $\hat{x}$ in the Heisenberg picture is given by 
\begin{equation}
\hat{x}(t)= \hat{x}(0)+{\hat{p}(0) \over m}\, t \:, \hspace{1cm}
t \geq 0\:. 
\label{HEIS}
\end{equation}
Let $\Delta x^{2}$ and $\Delta p^2$ denote the variances of the 
corresponding operators {\it i.e.}, 
$\Delta x^{2}=\langle \Delta \hat{x}^2 \rangle$, 
$\Delta p^{2}=\langle \Delta \hat{p}^2 \rangle$, with  
$\Delta \hat{x}=\hat{x}-\langle \hat{x} \rangle$, 
$\Delta \hat{p}=\hat{p}-\langle \hat{p} \rangle$, and 
$\langle \hat{x} \rangle$, $\langle \hat{p} \rangle$
being the average position and momentum values, respectively.
According to Eq. (\ref{HEIS}), the position variance after a time $t$
can be related to the initial uncertainties 
$\Delta x^2(0)$ and $\Delta p^{2}(0)$ by 
\begin{equation}
\Delta x^2(t)=\Delta x^2 (0)+ {\Delta p^{2}(0) \over m^{2}} \, t^{2} 
\ge \Delta x^{2}(0) \:,
\label{GROW}
\end{equation}
expressing the fact that the initial position variance increases under
free evolution. The SQL for position measurements coincides with the 
minimum value attainable by $\Delta x^2 (t)$ in (\ref{GROW}) when $t$ 
is regarded as the time interval between two successive, identical 
position measurements \cite{CAVES1}. 
The argument is based on a heuristic application of the Heisenberg 
uncertainty principle, whereby $\Delta x(0)$ is interpreted as the 
uncertainty due to the finite resolution of the first measurement, 
and $\Delta p(0)$ represents the momentum disturbance caused by the 
measurement (see, however, \cite{OZAWA2002} for a substantially more refined treatment). 
Because $\Delta x(0) \Delta p(0) \ge \hbar/2$, Eq. (\ref{GROW}) can be
rewritten as
\begin{equation}
\Delta x^2(t) \ge  \Delta x^2 (0) +
{\hbar^{2} t^{2} \over {4 m^{2} \Delta x^2(0)}} \:,
\label{BOUND}
\end{equation}
which, upon minimization with respect to $\Delta x^2 (0)$, implies
\begin{equation}
\Delta x^2(t) \ge {\hbar t \over m} = \Delta x^2_{SQL}(t)\:.
\label{SQLVal}
\end{equation}
Accordingly, the outcome of the second position measurement cannot be 
predicted with uncertainty smaller than the SQL value,  
$\Delta x (t)_{SQL} = ({\hbar t / m})^{1/2}$.

Yuen pointed out a major flaw in the above derivation, as by adding
in quadrature the initial position variance and the one propagated in time 
from the initial momentum variance as in Eq. (\ref{GROW}) it is implicitly 
assumed that position and momentum are initially uncorrelated. 
While this is correct, for instance, in the common case where the mass starts
in a minimum-uncertainty state, it is necessary in general
to relax this assumption, by replacing (\ref{GROW}) with the full expression: 
\begin{eqnarray}
\Delta x^2(t) & = & 
\langle \psi(0) | \hat{x}(t)^2 | \psi(0) \rangle
- \langle \psi(0) | \hat{x}(t) | \psi(0) \rangle^2  \nonumber \\
& = &
\Delta x^2(0)+ { \left\langle \{ \Delta \hat{x}(0), \Delta \hat{p}(0) 
\}\right\rangle \over m} 
\, {t}  +
{\Delta p^2 (0) \over m^2}\, t^{2}  \:,
\label{PARABOLA}
\end{eqnarray}
where $\{\,,\,\}$ denotes the anti-commutator operation.
Thus, the position variance is directly sensitive to the initial 
{\sl correlation coefficient}  $\Delta_{xp}^{2} (0)$, with 
\begin{equation}
\Delta_{xp}^{2} (0)=  \langle \{ \Delta \hat{x} (0),\Delta \hat{p} (0) \} 
\rangle = \langle \{ \hat{x}, \hat{p} \} \rangle 
-2 \langle \hat{x} \rangle \langle \hat{p}\rangle 
= -i \hbar + 2 \langle \hat{x} \hat{p}  \rangle 
-2 \langle \hat{x}\rangle \langle \hat{p} \rangle \:, 
\label{CORCOEF}
\end{equation}
upon explicitly using the canonical commutation rules for $\hat{x}, 
\hat{p}$ \cite{GOTTFRIED}.
Yuen also proposed an explicit class of quantum states, the 
so-called {\sl twisted coherent states}, where a {\sl negative} 
correlation term $\Delta_{xp}^{2} (0) <0$ is realized \cite{YUEN}. 
As a consequence of the parabolic dependence 
of $\Delta x^{2} (t)$ upon $t$ implied by (\ref{PARABOLA}), the 
initial position variance then narrows, attaining a minimum at an 
optimal time $\bar{t} > 0$. In addition, the variance remains below its 
initial value $\Delta x^2 (0)$ for the finite time interval 
$0 < t < 2 \bar{t}$, opening the possibility to evade the SQL  
for appropriate values of the parameters. 

It is interesting to note that the essence of Yuen's proposal relies 
on allowing {\sl complex} Gaussian states. In one dimension, 
for instance, consider a wave function of the form
\begin{equation}
\psi(x)= k \exp{(-\alpha x^{2})} \:, 
\label{complex}
\end{equation}  
with $\alpha \in {\cmplxs}$. By writing $\alpha=\alpha_{1}+i \alpha_{2}$, 
$\alpha_{1},\alpha_{2} \in {\rlxs}$, the normalization condition 
leads to $k=(2 \alpha_1/\pi)^{1/4}$ (up to an irrelevant phase factor). 
Because for such a state $\langle \hat{x} \rangle = 0$, 
$\langle \hat{p} \rangle =0$, and 
$\langle \hat{x}\hat{p} \rangle = i \hbar \alpha /{2 \alpha_1}$, the
corresponding correlation coefficient becomes
\begin{equation}
\Delta_{xp}^2 (0) =\hbar \, {\alpha - \alpha_1\over \alpha_1}
= - 2 \hbar \xi \:, 
\label{csi} 
\end{equation}
where the parameter $\xi=\alpha_2/2\alpha_1$ can be, in principle,
arbitrarily large. Note that the value $\xi=0$ corresponds to the standard
Gaussian state, while $\xi > 0$ implies a negative correlation term 
responsible for the desired non-monotonic behavior of $\Delta x^2 (t)$. 
Following Yuen's notation, the prototype wave function (\ref{complex})
can be generalized to the following family of states:
\begin{eqnarray}
\psi_Y (x)=\left( {m \omega \over {\pi \hbar |\mu-\nu|^2}} \right)^{1/4} 
\exp \left\{ -{m \omega \over {2 \hbar}} {1+i 2\xi \over{|\mu-\nu|^2}}
\, ( x - x_0 ) ^2
+ i {p_0 \over \hbar}  (x - x_0) \right \} \:, 
\label{YuenState}
\end{eqnarray}
where now $\langle \hat{x} \rangle_Y = x_0$, $\langle \hat{p}
\rangle_Y = p_0$ are real numbers, and the parameters $\mu,\nu$, and 
$\omega$ are related to the initial position and momentum variances 
and correlation as follows:
\begin{equation}
\Delta x^2_Y (0) ={\hbar \over 2 m \omega} |\mu - \nu|^{2} \:, 
\end{equation}
\begin{equation}
\Delta p^2_Y (0) ={\hbar m \omega \over 2} 
{1+4 \xi^{2} \over|\mu - \nu|^{2}} \:, 
\end{equation}
\begin{equation}
\Delta_{xp,Y}^2 (0) = -2 \hbar \xi \:. 
\label{xicorr}
\end{equation}
Note that $\alpha_1 = m \omega /2 \hbar | \mu - \nu |^2$ (compare 
Eq. (\ref{complex})), and the position variance evolves 
explicitly in time as:
\begin{equation}
\Delta x^{2}_Y (t)= {\hbar \over 2 m \omega} |\mu-\nu|^{2}
-{2\hbar \xi \over m} \,t + {\hbar \omega \over 2m} {1+4 \xi^{2} \over 
|\mu -\nu|^{2}} \, t^{2} \:. 
\label{yuen}
\end{equation}

Motivated by the comparison with the SQL, a natural strategy is to optimize 
the {\sl relative} variance obtained upon normalizing $\Delta x^2_Y (t)$
by the spreading expected at time $t$ for a state evolving at the SQL 
as given in (\ref{SQLVal}) {\it i.e.}, to minimize the quantity
\begin{equation}
\Lambda_Y (t) \equiv {\Delta x^2_Y (t) \over \Delta x^2_{SQL}(t)}= 
m {\Delta x^2_Y (t) \over {\hbar t}} \:.
\label{relative}
\end{equation}
By differentiating $\Lambda_Y(t)$, we get a minimum relative variance 
at a time 
\begin{equation}
t^{*}_Y= {|\mu-\nu|^{2} \over {\omega (1+4\xi^{2})^{1/2} } } \:.
\label{optimal}
\end{equation}
This criterion coincides, for large values of $\xi$, with the direct 
optimization of the absolute variance (\ref{yuen})
originally adopted in \cite{YUEN}:
\begin{equation}
\bar{t}_Y= {2 \xi |\mu-\nu|^{2} \over {\omega(1+4\xi^2)} } \:,
\end{equation}
thanks to the fact that the ratio $t^{*}_Y/\bar{t}_Y=(1+4\xi^2)^{1/2}/2\xi$ 
quickly approaches unity with increasing $\xi$. The relative criterion 
(\ref{relative}), however, 
has the advantage of allowing for a continuous comparison with the 
SQL value at a generic instant of time $t$. From equations (\ref{yuen}) and 
(\ref{optimal}),  the minimum relative variance is therefore
\begin{equation}
\Lambda_Y (t^*) = (1+4\xi^2)^{1/2}-2\xi \:,
\label{YRATIO}
\end{equation}
which quantifies the advantage in the use of a Yuen contractive state 
with respect to one with zero (or positive) correlations. As a
function of $\xi$, the ratio
in (\ref{YRATIO}) is monotonically decreasing in the contractivity 
region $\xi > 0$, a unit value corresponding to a classical state with 
$\xi=0$. While arbitrarily small values can be attained asymptotically,
a significant gain already appears for relatively small $\xi$ parameters;
for instance $\Lambda_Y(t^\ast_Y) \simeq 0.41$ for $\xi=1/2$.

As mentioned, Yuen's proposal sparked a lively debate on the meaning 
and conceptual consistency of the SQL for a free mass 
\cite{LYNCH,CAVES1,OZAWA,NI,BONDURANT}, and the discussion
still continues \cite{BRAGINSKY2001,OZAWA2002}. While 
various issues have been substantially elucidated, from an operational
perspective limited effort has been devoted to address possible ways 
for preparing Yuen states in the laboratory. To the best of our knowledge, 
the only steps in this direction have been taken by Storey and co-workers
\cite{STOREY}, who proposed a position measurement scheme for atoms
entering a strongly detuned standing light wave in a cavity, and 
by Vitali {\it et al.} \cite{VITALI0} in the context of stochastic cooling 
of macroscopic mirrors. Ultimately, one obstacle encountered in designing 
realizable preparation schemes for Yuen states is related to the difficulty 
of dealing with a complex $\alpha$. Thus, although very interesting in principle, this class 
of states seems not very palatable to experimentalists, motivating the search
for alternative families of contractive states.

\section{Contractive features of Schr\"odinger cat states} 

Following Yuen's original suggestion, position-contractive states 
for a free mass are defined by the property 
\begin{equation}
\left. {{d \over dt} \Delta x^2 (t)} \right|_{t=0} = 
\Delta_{xp}^2 (0) < 0 \:.
\label{DERIV}
\end{equation}
Because of the equation of motion (\ref{PARABOLA}), $\Delta x^2 (t)$ 
increases monotonically with time whenever $\Delta_{xp}^2 (0) \geq 0$.
Equivalently, in terms of the relative variance $\Lambda (t)$, one may 
verify that if $\Delta_{xp}^2 (0) \geq 0$, the inequality 
$\Lambda (t) \geq 1$ for all $t$ is also enforced as a consequence of  
the Heisenberg uncertainty principle, and vice versa. 
In particular, the condition (\ref{DERIV}) is never satisfied by a 
preparation in a classical Gaussian state. However, Yuen's class is not 
the only family of non-classical Gaussian states one may {\it a priori} 
consider. Another notable class capable, as we shall describe now, of 
fulfilling (\ref{DERIV}) is provided by {\sl macroscopically 
distinguishable} Gaussian states {\it i.e.}, Schr\"odinger cat states in the 
position variable. 

\subsection{Two-component cat states }

Let us consider, to begin with, a two-component Schr\"odinger cat state,
namely a coherent superposition of two one-dimensional Gaussian wave packets 
with inverse variance parameter $\alpha \in {\rlxs}^+$, localized around two positions $+x_{0}$ 
and $-x_{0}$, $x_0 >0$, and having relative amplitudes $k_+, k_- \in \cmplxs$:
\begin{equation}
\psi_{S2} (x)=N_2 \left\{k_+ \exp[-\alpha(x-x_{0})^{2}]+k_{-} 
\exp[-\alpha(x+x_{0})^{2}]\right\} \:.
\label{cat}
\end{equation}
Here, the normalization constant is given by 
\begin{equation}
|N_2|^2=(2\alpha/\pi)^{1/2} \left[
|k_+|^2+|k_-|^2+\exp(-2\alpha x_0^2)
(k_+^* k_-+k_-^*k_+) \right]^{-1} 
\:, \label{normaliz}
\end{equation}
and we have assumed that the two Gaussians have zero initial momentum. Note 
that, with this notation, each macroscopically distinguishable component 
in $|\psi_{S2} (x)|^2$ corresponds to a Gaussian probability
distribution with variance $1/4\alpha$.
After straightforward calculations we obtain the initial average values of 
position, momentum, and their product as follows:
\begin{equation}
\langle \hat{x} \rangle_{S2} = { |k_+|^2-|k_-|^2 \over 
{|k_+|^2+|k_-|^2+ \exp(-2\alpha x_0^2) (k_+^*k_-+k_-^*k_+)}} 
x_{0}\:, 
\label{xav}
\end{equation}
\begin{equation}
\langle \hat{p} \rangle_{S2} = 4\hbar \alpha { \text{Im} { (k_+ k_-^*)}
\over |k_+|^2-|k_-|^2 }\exp(-2\alpha x_0^2) \langle \hat{x} \rangle_{S2} \:, 
\label{pav}
\end{equation}
\begin{equation}
\langle \hat{x} \hat{p} \rangle_{S2} = i \hbar/2 \:. 
\label{xpprod}
\end{equation}
Thus, the correlation coefficient for a mass initially prepared 
in the state (\ref{cat}) is:
\begin{equation}
\Delta_{xp, S2}^{2}(0) = -2 \langle \hat{x} \rangle_{S2} 
\langle \hat{p} \rangle_{S2} = 
-8 \hbar \alpha { \text{Im}{ (k_+ k_-^*)} 
\over{|k_+|^2-|k_-|^2}} \exp(-2\alpha x_0^2) 
\langle \hat{x} \rangle_{S2}^2 \:. 
\label{sigmaxp}
\end{equation}
According to (\ref{sigmaxp}), two contractivity regions fulfilling 
(\ref{DERIV}) are then possible in principle:
\begin{eqnarray}
\text{(I)} \;\; \left\{ \, \begin{tabular}{c}
             $ |k_+|^2-|k_-|^2 >0 \:, $\\
             \text{Im}${ (k_+ k_-^*)} > 0\:,$ 
                      \end{tabular} \right. 
\hspace{1cm}\text{or} \hspace{1cm} 
\text{(II)} \;\; \left\{ \, \begin{tabular}{c}
             $ |k_+|^2-|k_-|^2 <0 \:, $\\
             \text{Im}${ (k_+ k_-^*)} < 0 \:.$ 
                      \end{tabular} \right.
\label{CONTRA}
\end{eqnarray}

This illustrates a different route for generating a non-zero correlation 
coefficient compared to Yuen states. In the latter case, regardless of 
whether $\langle \hat{x} \rangle_Y \langle \hat{p} \rangle_Y=0$ or 
not, a non-zero correlation $\Delta_{xp,Y}^{2}$ arises due to the 
asymmetric contribution from $\langle \hat{x} \hat{p} \rangle_Y$ and 
$\langle \hat{p} \hat{x} \rangle_Y$, enforcing $\langle \{ \hat{x},\hat{p} \} 
\rangle_Y \not = 0$. For Schr\"odinger cat states of the form (\ref{cat}), 
it is instead the initial anti-correlation between $\hat{x} $ and $\hat{p}$, 
$\langle \{ \hat{x},\hat{p} \} \rangle_{S2} = 0$, that induces contractivity 
for appropriate $k_+, k_-$ and $\langle \hat{x}\rangle_{S2} \not =0$. 
From a less formal viewpoint, in the case of the Yuen states the 
presence of an imaginary part in the quadratic coefficient $\alpha$ 
effectively correlates the position and momentum observables. For 
cat states as in (\ref{cat}), the correlation is obtained through 
the {\sl delocalized spatial structure}, which can enforce non-zero average 
values for both $\hat{x}$ and $\hat{p}$ as in (\ref{xav})-(\ref{pav}).

In order to evaluate the position variance at time $t$ based on 
(\ref{PARABOLA}), it is necessary to further calculate the second
moments $\Delta x^2_{S2} (0)$ and $\Delta p^2_{S2} (0)$ (see also the appendix). 
For convenience, we set $\alpha = 1/2\Delta_0^2$ henceforth, and introduce an 
adimensional time parameter as
\begin{equation}
\eta = {\hbar t \over m \Delta_0^2} \:.
\label{TIME}
\end{equation}
It is worth noting that while the values of $x_0$ and $\Delta_0$ 
separately affect the expectations $\langle \hat{x} \rangle_{S2}$, 
$\langle \hat{p} \rangle_{S2}$, the correlation coefficient (and hence 
contractivity) depends only upon their ratio, which we denote by 
$\delta= x_0/\Delta_0$. 
We can also assume, without loss of generality, that $k_+ \in {\rlxs}$. 
By letting $k_+=c_+$, $k_-=c_- e^{-i \theta}$, with $c_\pm, \theta \in 
\rlxs$ and $\theta$ determining the initial relative phase between the 
two Gaussian components, we finally reach the rewritten forms:
\begin{eqnarray}
\Delta x^2_{S2} (0) =  {\Delta_{0}^2 \over 2}+2 c_{+}c_{-} {{2 c_+ 
c_-+(c_+^2+c_-^2) e^{-\delta^2} \cos \theta} \over {D}}  \, x_0^2 \:,
\label{sigmax0}
\end{eqnarray}
\begin{eqnarray}
{\Delta p^2_{S2} (0) } = {\hbar^2 \over {\Delta_0^4}}
\left[ {\Delta_0^2 \over 2}-2 c_{+}c_{-} 
{2 c_+ c_- e^{-\delta^2} 
+(c_+^2+c_-^2) \cos \theta \over D} \, e^{-\delta^2} x_0^2 \right]  \:. 
\label{sigmap0}
\end{eqnarray}
In the above equations, $D=\left[c_+^2+c_-^2+2 c_{+} c_{-} 
e^{-\delta^2} \cos \theta\right]^2$. 
The complete expression for the position variance is then:
\begin{eqnarray}
\Delta x^2_{S2} (\eta) &=&  \left[{\Delta^2_0 \over 2}+2 c_{+}c_{-} 
{{2 c_+ c_-+(c_+^2+c_-^2)  e^{-\delta^2}  \cos \theta} 
\over {D}} \, x_0^2   \right] 
- \left[ 4 c_+ c_- {c_{+}^{2}-c_-^{2}\over D} \sin \theta 
e^{-\delta^2}  x_0^2 \right] \eta   \nonumber \\ 
&+ & \left[{\Delta_0^2 \over 2}-2 c_{+}c_{-} 
{2 c_+ c_- e^{-\delta^2} + (c_+^2+c_-^2) \cos \theta \over D} \,
e^{-\delta^2} x_0^2  \right] \eta^2 \:. 
\label{sigmaxt}
\end{eqnarray}

Thus, the variance may be written as $\Delta x^2_{S2} (\eta)={\tt A}+
{\tt B} \eta + {\tt C} \eta^2$, where the coefficients ${\tt A}, {\tt B}, 
{\tt C}$ of the relevant powers of $\eta$ can be inferred from the 
quantities in square brackets in Eq. (\ref{sigmaxt}). By construction, 
${\tt A}$ reproduces the initial position variance, while ${\tt B}$ and 
${\tt C}$ are responsible for the spreading from the initial value. The 
behavior for a single initial Gaussian state may be recovered by simply 
setting either $c_+$ or $c_-$ to zero. Note that, regardless of possible 
contractive features, the rate of spreading of $\Delta x^2_{S2} (\eta)$ 
may be slower than in the Gaussian case due to the interference term 
contained in ${\tt C}$. By normalizing $\Delta x^2_{S2} (\eta)$ to the reference value 
dictated by the SQL, $\Delta x^2_{SQL}(\eta)= \Delta_0^2 \eta$, we obtain:
\begin{equation}
\Lambda_{S2} (\eta) = {\Delta x^2_{S2} (\eta) \over \Delta x^2_{SQL}(\eta)} = 
{ {\tt A}_2 \over \eta}+{\tt B}_2+{\tt C}_2 \eta \:, 
\label{RHOCAT}
\end{equation}
where ${\tt A}_2 = {\tt A}/\Delta_0^2$ {\it etc}.
By minimizing $\Lambda_{S2}(\eta)$ with respect to $\eta$, 
an optimal time equal to 
${\eta}^* =({\tt A}_2/{\tt C}_2)^{1/2}$ is found, which is 
independent of the linear coefficient ${\tt B}_2$. The corresponding 
minimum value attained by $\Lambda_{S2}$ is 
\begin{equation}
\Lambda_{S2}({\eta}^* )= 
{\tt B}_2 + 2 ({\tt A}_2 {\tt C}_2)^{1/2} \:, 
\end{equation}
which can be further optimized with respect to the remaining parameters. 

To analyse in detail the behavior of $\Lambda_{S2}$, it is convenient to 
additionally introduce the ratio $\kappa={c_{+} /c_{-}}$, 
and rewrite the relative variance (\ref{sigmaxt}) as:
\begin{eqnarray}
\Lambda_{S2} (\eta, \kappa, \theta, \delta) &= &
\left[ {1\over 2} + 2 \kappa \delta^2 {2 \kappa+(1+\kappa^2) 
e^{-\delta^2} \cos \theta \over (1+\kappa^2 +2 \kappa e^{-\delta^{2}} 
\cos \theta)^2} \right] {1 \over \eta} + \left[
 4 \kappa \delta^2 {(1-\kappa^2) e^{-\delta^2} \sin \theta 
\over (1+\kappa^2+2 \kappa e^{-\delta^2} \cos \theta)^2} \right] 
\nonumber \\ 
& + & \left[ {1 \over 2} - 2\kappa \delta^2 {2 \kappa e^{-\delta^2}+
(1+\kappa^2)\cos \theta \over (1+\kappa^2+2 \kappa e^{-\delta^2} 
\cos \theta)^2} e^{-\delta^2} \right] \eta \:,
\label{final} 
\end{eqnarray}
where the dependence of $\Lambda_{S2}$ upon the relevant parameters has 
been made explicit, and the expressions in the square brackets now 
correspond to ${\tt A}_2,{\tt B}_2, {\tt C}_2$ of (\ref{RHOCAT}).
According to (\ref{final}), $\Lambda_{S2}$ is a complicated function of 
both the time $\eta$ and the various parameters $\kappa,\delta,\theta$ 
which characterize the initial state. In terms of the new parametrization,
the contractivity regions given in (\ref{CONTRA}) become: 
\begin{eqnarray}
\text{(I)} \;\; \left\{ \, \begin{tabular}{c}
             $ \kappa >1 \:, $\\
             $\sin \theta > 0 \:,$ 
                      \end{tabular} \right. 
\hspace{1cm}\text{or} \hspace{1cm} 
\text{(II)} \;\; \left\{ \, \begin{tabular}{c}
             $ \kappa < 1 \:, $\\
             $\sin \theta < 0 \:.$ 
                      \end{tabular} \right.
\label{CONTRABis}
\end{eqnarray}

Note that the dependence upon $\theta$ is periodic, thus we can limit 
the analysis to the range $0^\circ \leq \theta < 360^\circ$. In addition, 
$\Lambda_{S2}$ satisfies the following invariance property: 
\begin{equation}
\Lambda_{S2}(\eta, \kappa, \theta, \delta ) = 
\Lambda_{S2}(\eta, \kappa^{-1}, 360^\circ -\theta, \delta ) \:.
\label{INVAR}
\end{equation}
Thus, given the behavior of $\Lambda_{S2}$ in a single contractivity 
region, say (I), the behavior in the remaining region is the same upon
transforming $\kappa \mapsto \kappa^{-1}$ and $\theta \mapsto 
360^\circ -\theta$. Some qualitative insights into the behavior of the 
function (\ref{final}) can be obtained by inspection of the various terms. 
The contractive term, ${\tt B}_2$, assumes zero values at $\theta=0^\circ$ 
and $180^\circ$, namely, when the two distinguishable components of 
the cat add in phase or anti-phase, as well as when $\kappa=1$, which 
corresponds to an equally weighted superposition state. As a function of 
$\theta$ and $\delta$, $|{\tt B}_2|$ is large for $\theta \approx 90^\circ$ 
and $\delta$ of the order of unity.

\begin{figure}
{\epsfxsize=6in\centerline{\epsffile{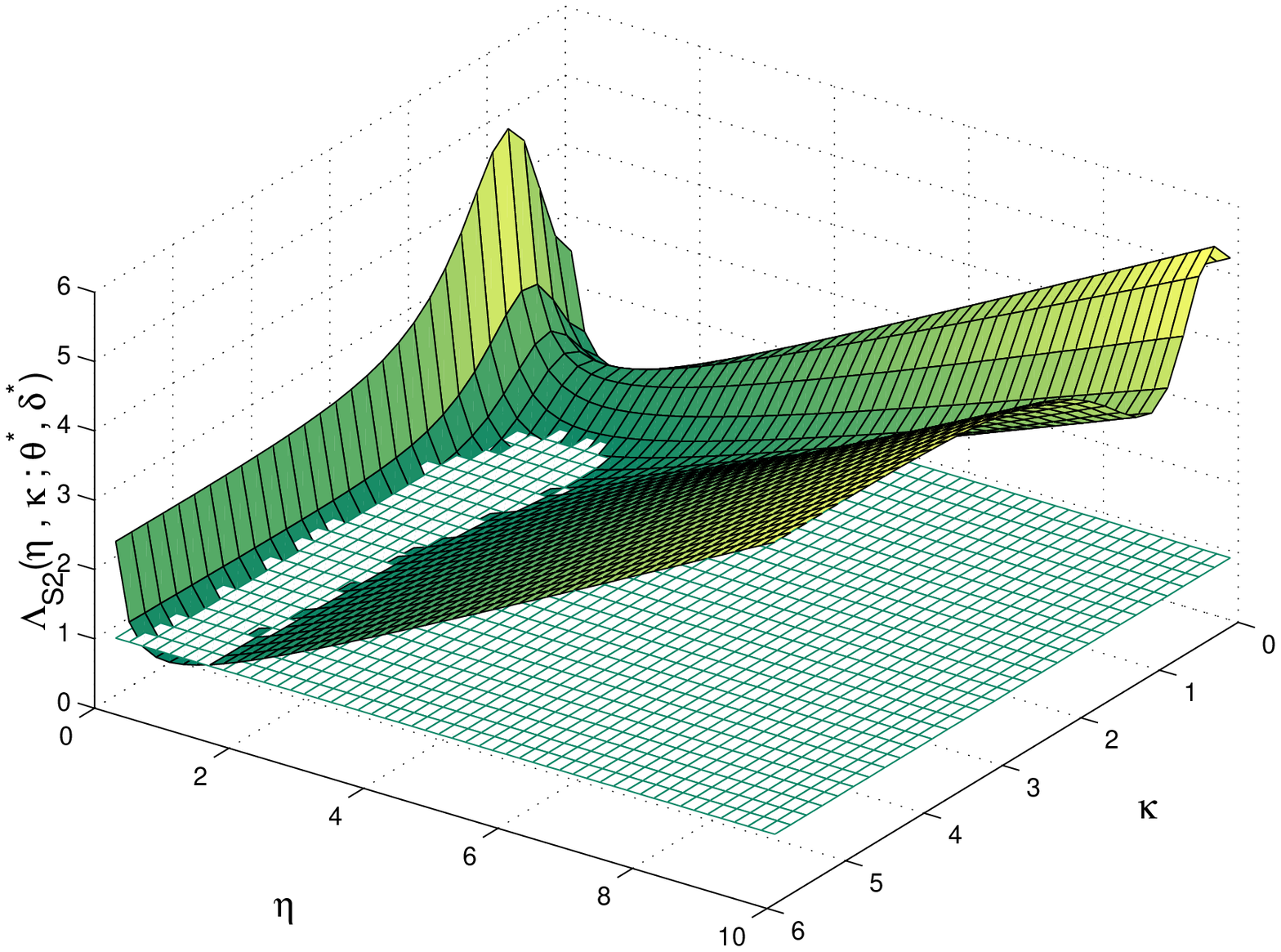}}}
\caption{\label{Fig1}
The dependence of the contractivity figure of merit $\Lambda_{S2}$ 
upon the effective time $\eta$ and $\kappa$ for optimal values 
$\theta^* \simeq 127^\circ$ and $\delta^*  \simeq 0.49$.
The intersection with the plane at constant $\Lambda_{S2}=1$ is also 
evidenced, showing the existence of an island where the variance 
of a Schr\"odinger cat state maintains values smaller than the SQL.} 
\end{figure}

\begin{figure}
{\epsfxsize=6in\centerline{\epsffile{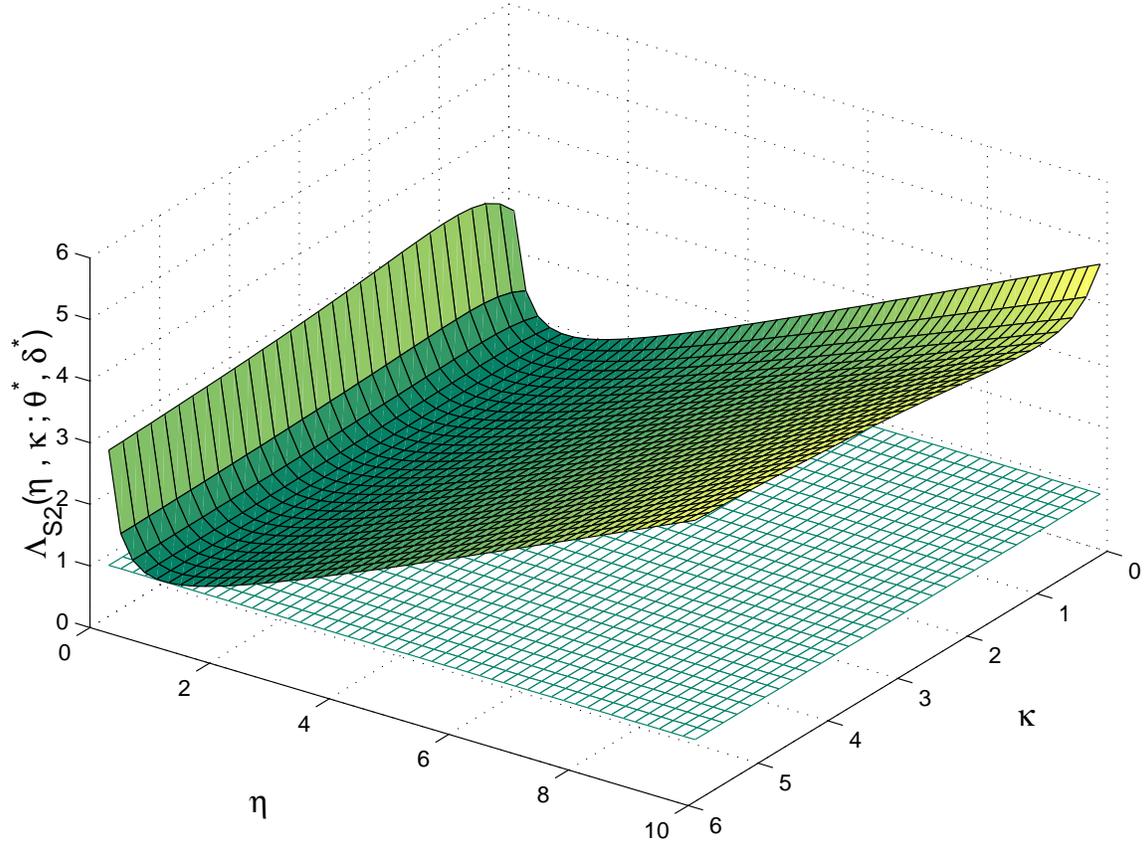}}}
\caption{\label{Fig2}
The dependence of the contractivity figure of merit $\Lambda_{S2}$ 
upon the effective time $\eta$ and $\kappa$ for $\theta=0^\circ$  
and $\delta^* \simeq 0.49$ as in Fig. (\ref{Fig1}).
No region with $\Lambda_{S2} <1$ is present in this case, 
therefore excluding any contractivity for the corresponding 
Schr\"odinger cat state.}
\end{figure}

\begin{figure}
{\epsfxsize=6in\centerline{\epsffile{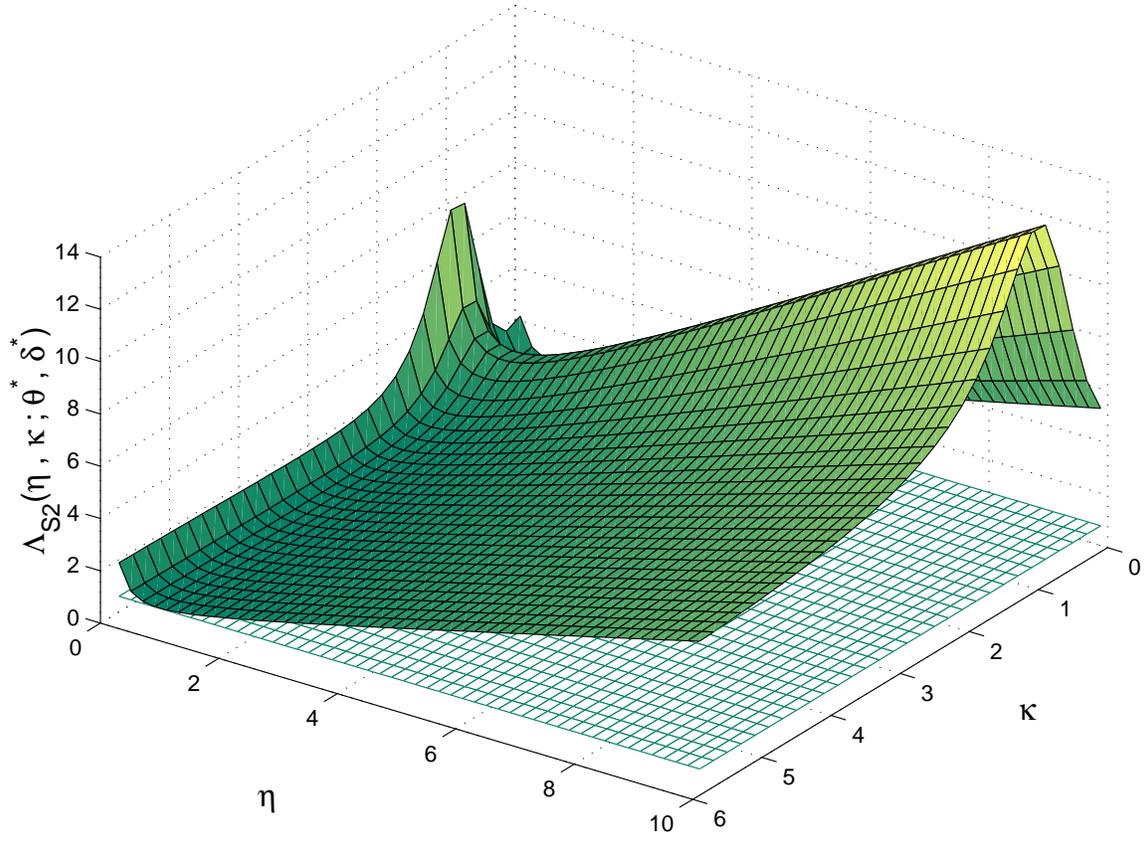}}}
\caption{\label{Fig3}
As figure (\ref{Fig2}), but for  $\theta=180^\circ$.   }
\end{figure}

By numerical analysis of Eq. (\ref{final}), the figure of merit for
contractivity $\Lambda_{S2}$ is found to attain its minimum 
\begin{equation}
\Lambda_{S2}^{min}(\eta^*,\kappa^*, \theta^*, \delta^*) \simeq 0.757 \:, 
\label{MINIMUM}
\end{equation}
for optimal parameter values $\eta^* \simeq 1.105$, $\kappa^* 
\simeq 2.26$, $\theta^* \simeq 127^\circ$, and $\delta^* \simeq 0.49$.
The dependence of $\Lambda_{S2}$ upon the effective time $\eta$ and 
$\kappa$ is displayed in Fig. (\ref{Fig1}) for $\theta=\theta^*$ and 
$\delta=\delta^*$. The section in the $(\eta,\kappa)$ plane resulting 
from a cut with the plane $\Lambda_{S2}=1$ is also depicted. The plot 
evidences a region where $\Lambda_{S2}$ stays below 1. For 
comparison, in Figs. (\ref{Fig2}) and (\ref{Fig3}), the corresponding 
dependence is shown for a state which only differs in the initial 
relative phase,  $\theta=0^\circ$ and $180^\circ$, respectively. 
The region $\Lambda_{S2} <1$ is not entered for such states. 

\begin{figure}
{\epsfxsize=6in\centerline{\epsffile{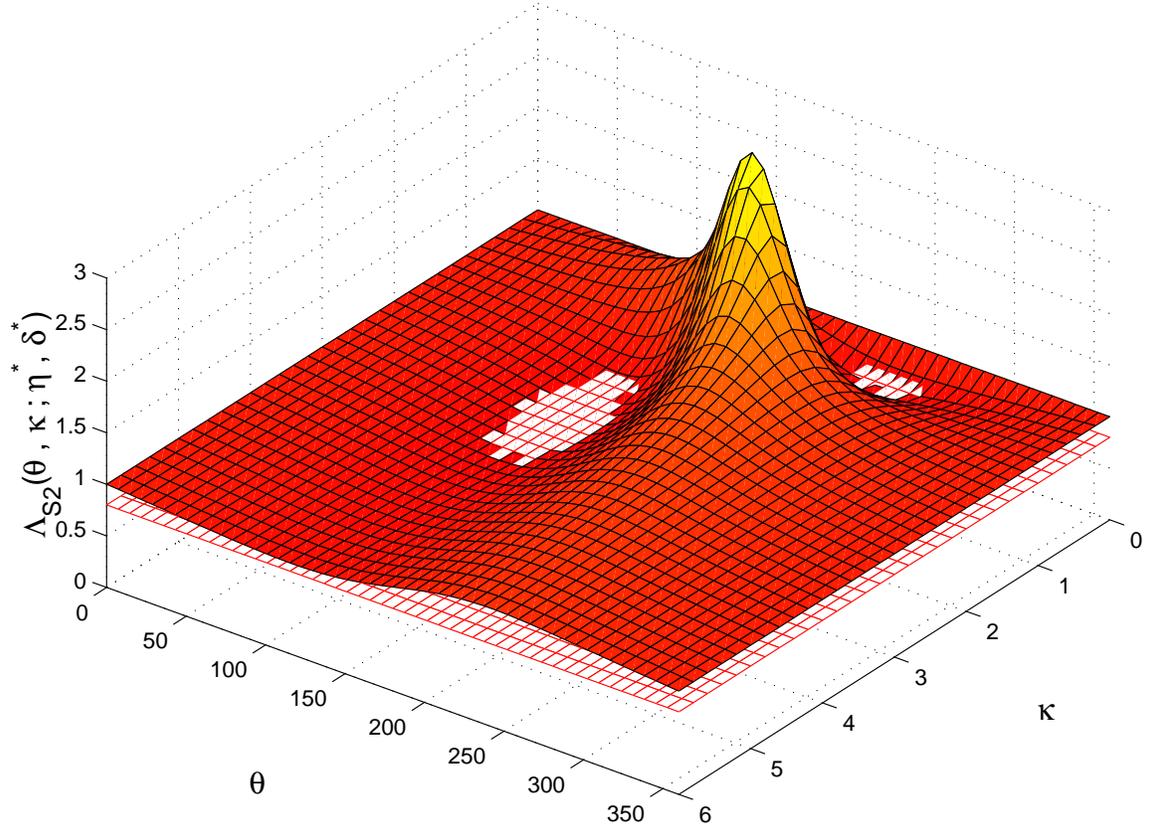}}}
\caption{\label{Fig4}
The dependence of the contractivity figure of merit $\Lambda_{S2}$ upon 
$\theta$ and $\kappa$ for optimal values $\eta^* \simeq 1.105$ 
and $\delta^* \simeq 0.49$. The two islands result from the section
at constant $\Lambda_{S2}=0.8$, and are connected via the relationship 
given in equation (\ref{INVAR}). }
\end{figure}

\begin{figure}
{\epsfxsize=6in\centerline{\epsffile{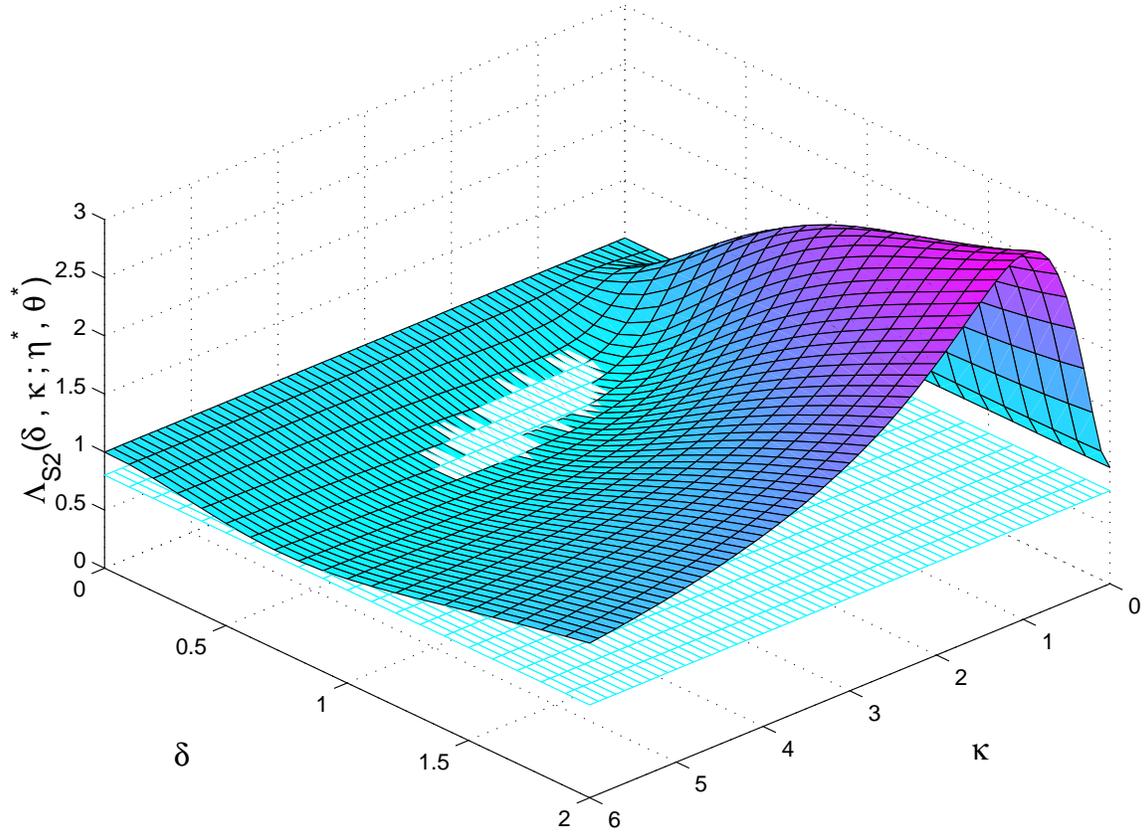}}}
\caption{\label{Fig5}
The dependence of the contractivity figure of merit $\Lambda_{S2}$ upon 
$\kappa$ and $\delta$ for optimal values $\eta^* \simeq 1.105$ 
and $\theta^* \simeq 127^\circ$. As in figure (\ref{Fig4}), the 
intersection with the plane $\Lambda_{S2}=0.8$ is also displayed for
reference. }
\end{figure}

By keeping the effective time fixed at the value $\eta=\eta^*$, 
one can focus on some other dependencies as illustrated in Figs. 
(\ref{Fig4}) and (\ref{Fig5}). In particular, Fig. (\ref{Fig4}) 
shows the dependence upon $\theta$ and $\kappa$. 
In this case, a reference contour plot is chosen at constant 
$\Lambda_{S2}=0.8$.  The two contractivity islands characterized by 
(\ref{CONTRABis}) are clearly visible, and related to each other as 
in (\ref{INVAR}). In Fig. (\ref{Fig5}), a similar plot depicts 
the dependences upon $\delta$ and $\kappa$.

\subsection{Generalizations: Three-component cat states }

The above analysis shows that contractive quantum states, in the 
original Yuen's spirit, can be engineered through an appropriate choice
of parameters in the class of cat states described by (\ref{cat}). 
Several variants might be worth exploring in principle. For instance, 
cat states arising from the superposition of two Gaussians differing in 
their values of the parameters $x_0$ or $\Delta_0$, or possessing non-zero 
initial momenta could be examined, as well as superpositions of 
non-Gaussian wave-packets.  

Another direction for generalizations is to consider {\sl multi-component 
cat states}, for which contractivity could be enhanced via coherent interference 
effects between different pairs. We illustrate this possibility by focusing 
on the simplest generalization of the class (\ref{cat}), leading to a family 
of cat states with three macroscopically distinguishable components. 
Thus, let us consider a wave-function of the form
\begin{equation}
\psi_{S3} (x)= N_3 \left\{
k_{+} \exp[-(x-x_0)^2/2\Delta_0^2]+
k_{0} \exp[-x^2/2\Delta_0^2]+
k_{-} \exp[-(x+x_0)^2/2\Delta_0^2]
\right\} \:,
\label{cat3}
\end{equation}
where the normalization constant is now given by 
\begin{eqnarray}
|N_3|^2 & = & (\pi \Delta_0^2)^{-1/2} 
\left[ 
|k_+|^2+|k_0|^2+|k_-|^2+(k_+^* k_-+k_+ k_-^*) 
e^{-\delta^2} 
\right. \nonumber \\  & + & \left. 
(k_+^* k_0+k_+ k_0^*) e^{-\delta^2/4 }+
(k_-^* k_0+k_- k_0^*) e^{-\delta^2/4} 
\right]^{-1} \:.
\label{normaliz3}
\end{eqnarray}
Here, $\delta=x_0/\Delta_0$ as before, and we also continue to assume that 
the three Gaussians have zero initial momentum. Because a global phase factor 
is irrelevant, the state (\ref{cat3}) is parametrized by five real parameters 
describing the relative amplitudes and phases, in addition to $x_0$ and 
$\Delta_0$, which are taken as before to be positive. 
Similar to the two-component cat case, we arbitrarily choose one of the $k$
coefficients to be real, thus setting $k_0=c_0$, $k_+= c_+ e^{i \theta_+}$, 
$k_-= c_- e^{i \theta_-}$, with $c_0,c_\pm, \theta_\pm \in \rlxs$. With these
conventions, the results for two-component cat states of the previous section
can be recovered by letting $c_0=0, \theta_+=0$, and $\theta_-= -\theta$. 
We also define
\begin{equation}
K =c_+^2+c_0^2+c_-^2+2c_0(c_+ \cos \theta_++c_- \cos \theta_-)e^{-\delta^2/4}+
2 c_+ c_- \cos(\theta_+-\theta_-) e^{-\delta^2} \:. 
\end{equation}
The expectations of the position and momentum observables on the state
(\ref{cat3}) take a more complicated form than in (\ref{xav})-(\ref{pav}):
\begin{equation}
\langle \hat{x} \rangle_{S3} ={1 \over K} 
\left[ c_+^2-c_-^2+c_0(c_+ \cos \theta_+-c_-\cos \theta_-)
e^{-\delta^2/4}  \right] x_0 \:, 
\end{equation}
\begin{equation}
\langle \hat{p} \rangle_{S3} ={\hbar \over \Delta_0^2} 
{{c_0 (c_+ \sin \theta_+-c_- \sin \theta_-) e^{-\delta^2/4}+2 c_+ c_- 
\sin(\theta_+-\theta_-) e^{-\delta^2}}\over
{c_+^2-c_-^2+c_0(c_+ \cos \theta_+-c_-\cos \theta_-) e^{-\delta^2/4}}}\,
\langle \hat{x} \rangle_{S3} \:.
\end{equation}
Similarly, one also finds 
\begin{eqnarray} 
\langle \hat{x} \hat{p} \rangle_{S3} & = & 
i {\hbar \over 2 K} \left[ c_+^2+c_0^2+c_-^2+2c_0(c_+ \cos \theta_+
+c_- \cos \theta_-)e^{-\delta^2} 
\right. \nonumber \\  & + & \left.
2 c_+ c_- \cos(\theta_+-\theta_-) e^{-\delta^2} - 2 i c_0 \delta^2 
(c_+ \sin \theta_++c_- \sin \theta_-) e^{-\delta^2/4} \right] \:.
\end{eqnarray}
Thus, unlike in the ordinary cat case leading to Eq. (\ref{xpprod}), 
$\langle \hat{x} \hat{p} \rangle_{S3}$ acquires in general a 
non-vanishing real component. While one can verify that 
Im$(\langle \hat{x} \hat{p} \rangle_{S3})=i \hbar/2$ and hence that a real
value of the correlation coefficient is correctly enforced, the full
expression for $\Delta_{xp,S3}^2 (0)$ is less transparent than in the 
two-component case. In formal analogy to Eq. (\ref{xicorr}), we may 
write 
\begin{equation} 
\Delta_{xp, S3}^2 (0)= 2 \Big( \text{Re}( \langle \hat{x} \hat{p} 
\rangle_{S3} ) - \langle \hat{x} \rangle_{S3} \langle \hat{p} 
\rangle_{S3} \Big)  = -2 \hbar \zeta  \:,
\label{zetacorr}
\end{equation}
where
\begin{eqnarray}
\zeta  =  {\delta^2 \over K^2} \Big[\hspace*{-1mm}
& - & \hspace*{-1mm} c_0 (c_+ \sin \theta_++c_- \sin \theta_-) 
K e^{-\delta^2/4} +
\Big(c_0(c_+ \sin \theta_+-c_- \sin \theta_-) e^{-\delta^2/4} 
\nonumber \\
\hspace*{-1mm} & + & \hspace*{-1mm}
 2 c_+ c_- \sin(\theta_+-\theta_-) e^{-\delta^2}\Big) 
\Big(c_+^2-c_-^2+c_0(c_+ \cos\theta_+ -c_- \cos \theta_-) e^{-\delta^2/4}
\Big) \Big] \:. 
\end{eqnarray}
The above function satisfies the property that
\begin{equation}
\zeta (c_+, c_-, \theta_+, \theta_-) = \zeta (c_-, c_+, \theta_-, \theta_+) 
\:, \end{equation}
{\it i.e.}, it is invariant under the exchange $k_+ \mapsto k_-$. 
Thus, one may expect $\zeta$ to be extremal for $k_+=k_-$, or 
$c_+ = c_-$ and $\theta_+ = \theta_-$. The fact that under these conditions 
contractive behavior is possible for a cat state of the form (\ref{cat3})
is easily verified by noting that $ \langle \hat{x} \rangle_{S3} =0$, 
$ \langle \hat{p} \rangle_{S3} =0$, and 
\begin{equation}
\zeta_{ [c_+ = c_-; \theta_+ = \theta_-]} = 
-2 \, {\delta^2 \over K}\, c_0 c_+ \sin \theta_+ e^{-\delta^2/4} \:.   
\label{class3}
\end{equation}
Thus, $\zeta >0$ for $\theta_+ \in (180^\circ, 360^\circ)$, which in turn 
corresponds to a family of cat states with a negative correlation term. 
Note that this is in contrast with the two-component case, where a 
non-vanishing expectation of $\hat{x}$ (hence $\hat{p}$) was found to
be necessary for contractivity. 

\begin{figure}
{\epsfxsize=5in\centerline{\epsffile{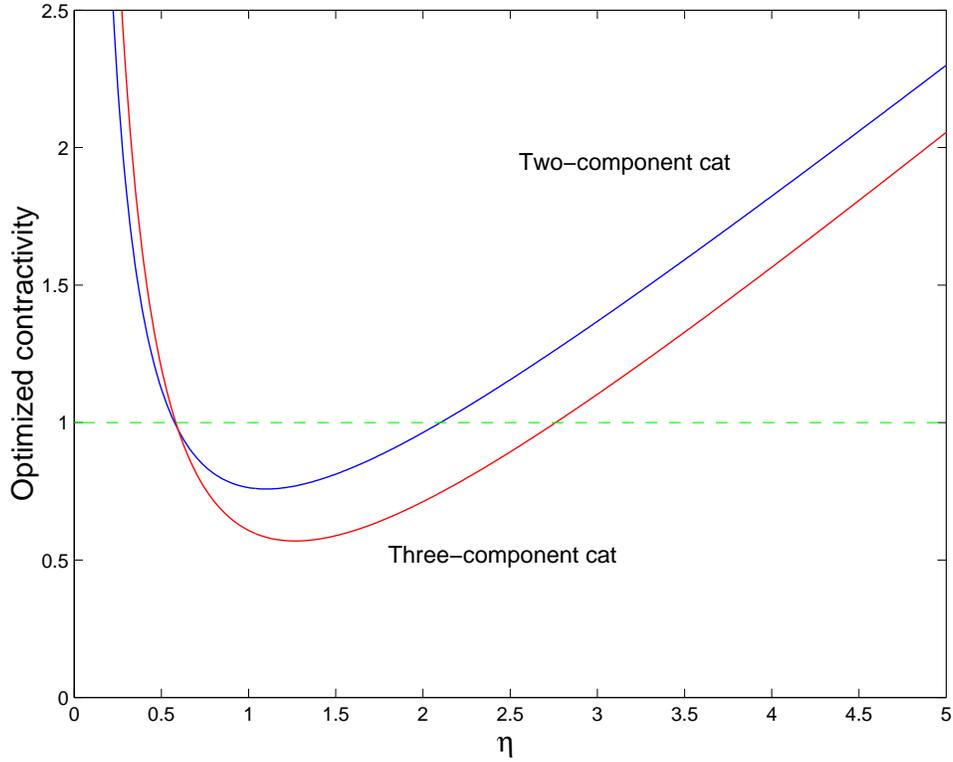}}}
\caption{\label{Fig6} 
Comparison of the optimized contractivity figures of merit $\Lambda^*_{S2}$ 
(solid blue) and $\Lambda^*_{S3}$ (solid red) versus rescaled time for two- 
and three-component cat states as in equations (\ref{cat}) and (\ref{cat3}), 
respectively. Except for the time, all the parameters are set to their optimal 
values as found upon minimization of the appropriate relative variance 
function. Note that lower values are attained by $\Lambda_{S3}^*$ over a wider 
time interval than $\Lambda_{S2}^*$. For reference, the unity value of a 
non-contractive state (dashed green) is also depicted.}
\end{figure}

The complete calculation of the position variance $\Delta x^2_{S3}(\eta)$ 
as a function of the rescaled time is slightly simpler in the Schr\"{o}dinger 
representation in this case. The essential steps and final equations are 
quoted in the appendix. Working as before in terms of a relative variance, 
\begin{equation}
\Lambda_{S3}(\eta) =  {\Delta x^2_{S3} (\eta) 
\over \Delta x^2_{SQL}(\eta)} = { {\tt A}_3 \over \eta}+{\tt B}_3+
{\tt C}_3  \eta \:, 
\label{RHOCAT3}
\end{equation}
explicit expressions for the coefficients ${\tt A}_3, {\tt B}_3,  
{\tt C}_3$ may be derived from the appendix. The behavior of 
$\Lambda_{S3}$ has been analyzed numerically in the parameter space 
corresponding to the set $\eta, \kappa_+, \kappa_-, \theta_+, 
\theta_-, \delta$, with $\kappa_+ = { c_+ / c_-}$ and $\kappa_- = 
{ c_0 / c_-}$. 
In this case, the global optimization leads to a minimum value
\begin{equation}
\Lambda_{S3}^{min}(\eta^*,\kappa^*_+, \kappa^*_-, \theta^*_+, 
\theta^*_-, \delta^*) \simeq 0.569  \:, 
\label{MINIMUM3}
\end{equation}
for optimal parameter values $\eta^* \simeq 1.270$, 
$\kappa^*_+ \simeq 1.00$, $\kappa^*_- \simeq 2.38$, 
$\theta^*_+ \simeq 249^\circ$, $\theta^*_- \simeq \theta^*_+$, 
and $\delta^* \simeq 1.21$.  The fact that $\kappa^*_+ \simeq 1.00$ and
$\theta^*_- \simeq \theta^*_+$ within the numerical accuracy indicates that 
the minimum of $\Lambda_{S3}$ is indeed attained in the parameter regime 
expected from (\ref{class3}). 
A pictorial comparison between the behavior of the optimized contractivity 
figures of merit $\Lambda^*_{S2}$ and $\Lambda^*_{S3}$ as a function of 
$\eta$ for two- and three-component cat states is displayed in figure 
(\ref{Fig6}). Is it worth noting that not only 
$\Lambda^*_{S3} < \Lambda^*_{S2}$ for $\eta \gtrsim 0.5$, but also the region 
where values lower than one are achieved appreciably widens for three-component 
contractive cat states. 

\begin{figure}
{\epsfxsize=5in\centerline{\epsffile{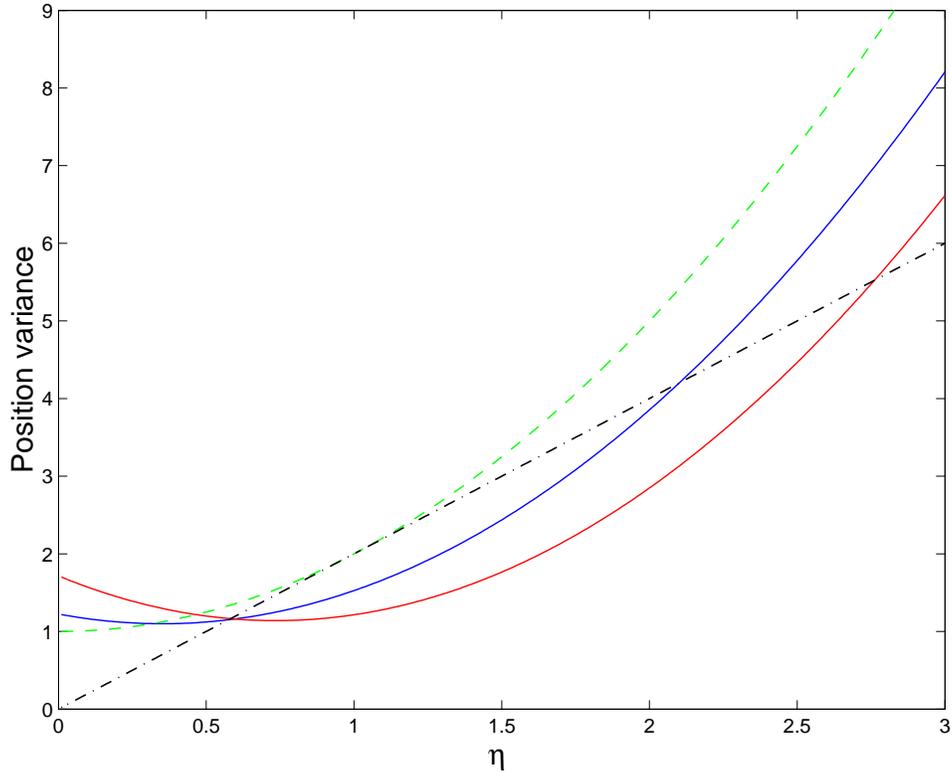}}}
\caption{\label{Fig7} 
The dependence of the position variance $\Delta x^2_{S2}$ (solid blue) and 
$\Delta^2_{S3}$ (solid red) versus rescaled time for two- and three-component 
cat states. The variance is evaluated in units of $\Delta_0^2/2$, which is the 
width of the probability distribution for each single Gaussian component of the 
cat. Except for the time, all the remaining parameters are set to the values maximizing 
contractivity as given in equations (\ref{MINIMUM}) and (\ref{MINIMUM3}), respectively. 
For comparison, the variance of a single-Gaussian state evolving as in 
equation (\ref{GAUSS}) from an initial variance $\Delta x^2 (0)= \Delta_0^2/2 =1/2$ 
(dashed green) is also depicted in the same units.  The SQL behavior corresponds, 
in these units, to a line (dash-dotted black) with slope 2 which is tangent to 
the Gaussian curve at $\eta=1$ and intersects the contractive curves as shown. }
\end{figure}

For an additional comparison between contractive and non-contractive
behavior, it is also instructive to directly examine the time dependence of 
the absolute variance for representative two-, three-component cat states, and 
for a Gaussian wave-packet with the same width parameter $\Delta_0$. The 
results are shown in figure(\ref{Fig7}) for maximally contractive cat states. 
The monotonic increase of the variance for the Gaussian case,
\begin{equation}
\Delta x^2_G (\eta) = {\Delta^2_0 \over 2} \, (1 +\eta^2)\:,
\label{GAUSS}
\end{equation}
is clearly visible. The SQL behavior of Eq. (\ref{SQLVal}), which translates 
into $\Delta^2_{SQL}(\eta) = \Delta^2_0 \eta$, is reached at time $\eta=1$ 
in the non-contractive Gaussian case of (\ref{GAUSS}). 
The contractivity intervals for two- and three-component cat states 
are determined by the appropriate intersections
of the variance curves $\Lambda_{S2}$, $\Lambda_{S3}$ with the SQL-line.  
Viewed in this way, the advantage of contractive states also manifests itself 
in the ability to preserve variances comparable to the initial Gaussian 
value at later times. In addition, the stronger contractivity of a 
three-component cat state is evidenced by the higher initial slope, 
implying both a deeper excursion below the SQL value and a longer 
contractivity interval.

To summarize, the larger set of parameters available in more complex classes 
of Schr\"{o}dinger cat states is capable of supporting stronger and better 
contractive properties. While further investigation is needed to precisely
characterize all the possible configurations, the fact that states with 
increased complexity may allow for a richer ``self-interference'' pattern 
(and hereby enhanced contractivity) is likely to occur beyond the specific 
case examined here. In particular, this is also somewhat reminiscent of 
the recent findings in \cite{ZUREK2001}, where cat states such as 
``compass-like'' superpositions of four Gaussians are able to probe 
sub-Planck structures in the phase space.

\section{ Implications of contractive cat states }

The existence of contractive cat states immediately evokes, as in Yuen's 
original proposal, the potential for breaching the SQL of a free mass.
However, a full assessment of such a potential requires a proper formulation 
of both the SQL itself and of various measurement-theoretical concepts such 
as quantum precision, resolution, and uncertainty. While addressing these
issues goes beyond the scope of the present work, as far as the implications 
for the SQL are concerned we limit ourselves to a few preparatory remarks. 

As emphasized by Ozawa \cite{OZAWA}, because the SQL applies to {\sl 
repeated} quantum measurements, one essential ingredient is the ability 
to perform a measurement using a {\sl single} measuring apparatus, the 
reading of which provides accurate information about the position of 
the free mass and simultaneously prepares it for the next position 
measurement, leaving it in a contractive state. A class of measurement 
schemes well suited for this task is offered by so-called {\sl Gordon-Louisell 
(GL) position measurements} \cite{GORDON}. The latter are naturally described 
within the general formalism of quantum operations \cite{KRAUS,CHUANG}. 
Let $a \in \rlxs$ be a real parameter, to be interpreted as the possible 
outcome of a position readout for a one-dimensional system, and let 
$\{ |a\rangle \}$ denote the complete set of position eigenstates. 
If $\{ \Psi_a\}$ is a family of normalized wave functions indexed by $a$, 
and $\rho(0^-)$ is the state of the system prior to the measurement, 
the state change and measurement statistics corresponding to a GL position 
measurement $\{ |\Psi_a\rangle\langle a|\}$ are described by the following 
operation measure and effect measure, respectively:
\begin{equation}
\rho(0^+) %= {\cal L}(I)[\rho(0^-)] 
= \int_I da\, A_a \rho(0^-) A_a^\dagger \:, 
\hspace{5mm} A_a=|\Psi_a\rangle\langle a|\:, 
A_a^\dagger=|a\rangle\langle \Psi_a|\:, 
\label{GLop}
\end{equation}
\begin{equation}
\text{Prob}(a \in I | \rho(0^-)) = \text{Trace}\Big[
\int_I da \,A_a^\dagger  A_a \rho(0^-) \Big]= 
\text{Trace}\Big[ \int_I da\, |a\rangle\langle a| \rho(0^-) \Big] \:, 
\label{GLeff}
\end{equation}
for all Borel sets $I \subseteq \rlxs$. The distinctive feature of GL
position measurement is that, according to (\ref{GLop}), the posterior 
state $\rho(0^+)$ of the system is determined by $\Psi_a$ {\sl 
regardless of the prior state} $\rho(0^-)$, the latter determining 
the outcome probabilities according to (\ref{GLeff}). Because GL 
measurements are described by completely-positive quantum maps, every 
GL measurement is physically realizable in principle \cite{OZAWA}. 

GL contractive position measurements were originally invoked by Yuen 
\cite{YUEN}. Given contractive states of the form (\ref{YuenState}), 
the relevant family $\{ \Psi_a\}_Y$ may be constructed by fixing the 
parameters $\mu, \nu, \omega$, and by identifying 
$x_0 =\langle \hat{x} \rangle_Y = a$, for all real $a$. 
If we now look, for instance, at two-component cat states as in (\ref{cat}),
then a contractive state satisfying that $\langle \hat{x} \rangle_{S2} = a$ 
can be found except for the (zero-measure) set $I =\{ a=0 \}$. Let 
$\beta >1$ a positive number. Then one may define a family $\{ \Psi_a\}_{S2}$ 
as follows:
\begin{eqnarray}
\{ \Psi_a \}_{S2} =  \left\{ \, \begin{tabular}{ll}
             $ \{ | \kappa_> =\sqrt{(\beta-1)/(\beta+1)} ; \theta =270^\circ; 
 x_0 = \:\beta a , \Delta_0 = \:\beta a /\delta    \rangle   \}     $ & 
\hspace{3mm} $a >0 \:,$ \\
             $ \{ | \kappa_< =\sqrt{(\beta+1)/(\beta-1)} ; \theta =90^\circ; 
 x_0 = -\beta a , \Delta_0 = -\beta a/
              \delta \rangle   \}    $ & 
\hspace{3mm} $a < 0 \:.$ 
             \end{tabular} \right. 
\label{BETA}
\end{eqnarray}
For fixed $|a|$, the two corresponding states in (\ref{BETA}) have the 
same degree of contractivity thanks to the relationship (\ref{INVAR}). 
By choosing $\delta = \delta^*$ and adjusting $\beta$ in such a way that 
$\kappa_< =\kappa^*, \kappa_>=1/\kappa^*$, contractivity can then be 
brought close to optimality. Establishing whether a family of contractive 
cat states such as (\ref{BETA}) actually supports a well-defined GL 
position measurement, and whether the latter may succeed in breaking the SQL, 
are interesting problems {\it per se}, which would call for an analysis 
along the lines of \cite{OZAWA}. Once a successful strategy is identified 
in principle, an additional issue would be looking for an explicit interaction 
Hamiltonians implementing the desired contractive measurement
scheme. For Yuen states, such a constructive problem was solved by Ozawa \cite{OZAWA}. 
It is interesting to note that the proposed measurement model explicitly 
requires the initialization of the meter in a contractive state, and 
contractivity is subsequently transferred to the mass via a controlled 
interaction. In this context, cat states might then turn advantageous if 
preparation procedures simpler than the ones generating Yuen states could 
be exploited to appropriately initialize the meter.

Independently of the SQL motivation, the possibility of experimentally 
demonstrating and characterizing contractive behavior for a free massive 
particle is both interesting as a fundamental quantum property, and 
for its possible implications in high-precision {\sl single} quantum 
measurements. While being certainly challenging with present capabilities, 
accomplishing this goal could largely benefit from the extensive efforts 
which are under way to realize Schr\"odinger cat states in diverse physical 
systems. Starting from the pioneering proposals by Yurke and Stoler \cite{YURKE,YURKE1}
for the generation of optical cat states, and related 
following proposals \cite{SCHLEICH,SONG}, such efforts have led 
to the successful creation and manipulation of 
macroscopically distinguishable photon states in the cavity QED setting 
\cite{HAROCHE}, and have culminated in the generation of a mesoscopic 
Schr\"odinger cat for a trapped Be ion \cite{MONROE}. 
Also, the creation of macroscopically distinguishable states has been 
recently reported for superconducting circuits \cite{ROUSE,NAKAMURA,FRIEDMAN},
and proposals for entanglement of trapped electrons have been recently
formulated \cite{MASSINI}.  
Interesting proposals exist for generating Schr\"odinger cat 
states of mechanical systems via controlled entanglement with 
microscopic degrees of freedom, including electron pairs in Cooper 
boxes \cite{ARMOUR}, single photons in beam-splitter configurations 
\cite{MARSHALL}, as well as radiation-pressure controlled 
mirrors \cite{FABRE,MANCINI,HEIDMANN,MANCINI1,BOSE,ZHENG}. 
Finally, recent advances in ultracold atom physics open 
new perspectives for the possibility of creating Schr\"odinger cat states 
with ultracold atomic ensembles or gaseous Bose-Einstein condensates 
\cite{CIRAC,GORDONF}. 

Devices based on trapped ions or cold atoms seem especially promising
for demonstrating and exploiting contractivity properties. In this case, 
if one of the strategies mentioned above actually succeeds for engineering
a contractive Schr\"odinger cat state, trapping potentials could be 
switched off and the mass (ion or atomic cloud) be released to free 
evolution for a given time interval. The position variance could then be 
inferred from the spatial pattern of the fluorescent light emitted under 
the action of a light probe beam, and by repeating the experiment for 
different times of flight the relevant time dependence could be 
reconstructed. Besides demonstration, contractivity might find useful 
applications in high-precision atomic physics experiments. For instance, 
one possibility worth exploring in principle is trying to improve 
the accuracy of atomic-fountain clocks through proper refocusing of the 
atomic cloud on the fluorescence detection area, as this could translate 
into a decrease of the absolute uncertainty on the atomic populations 
\cite{SANTARELLI}. A general observation which may be relevant in this
context is that, because the behavior of the position variance
would be unchanged for a mass subject to a linear potential, contractive
states for a free mass would still exhibit contractivity under a 
{\sl uniform gravitational field} (see also \cite{MICROCAT} for a related 
discussion). Thus, contractivity could in principle be exploited both
in the vertical and horizontal directions. 

In spite of these promising avenues, a potentially serious concern 
arises from the fact that Schr\"odinger cat states may be especially 
fragile against the decoherence effects caused by the coupling to their 
surrounding environment, a feature which is regarded as crucial to 
understanding the quantum-to-classical transition \cite{ZUREK,GIULINI}. 
Under the assumption that the environment may be schematized as a 
harmonic bosonic bath, exact analyses are possible by adapting the 
results available for the quantum Brownian motion model 
\cite{GIULINI,PAZ}, or by directly resorting to results already
obtained for a free damped quantum particle \cite{JUNG,HAKIM}.
 However, it is also worth stressing that the decoherence 
behavior may be quite sensitive to various details of the 
system-environment coupling as well as environment properties, 
and there is hope that on one hand estimates based on general models 
might be in some cases overly pessimistic \cite{TESSIERI}, and on the
other hand decoherence effects might be effectively counteracted. 
In particular, stabilization procedure for cat states could be in 
principle designed based on the basis of both active-control schemes 
for the system \cite{VIOLA,TOMBESI} and symmetrization schemes 
for the environment \cite{DALVIT}.

\section{Conclusions} 

We have shown that Schr\"{o}dinger cat states provide a new class of 
states able to support contractive features in the sense originally 
proposed by Yuen in \cite{YUEN}. Yuen contractive states lead in 
principle to an unlimited gain with respect to the SQL for a free 
particle but, apart from a proposal in the atomic physics context 
\cite{STOREY}, it is not clear how to prepare such states in practice. 
Contractive Schr\"{o}dinger cat states, on the other hand, may enable 
one to  capitalize on the intense effort which is ongoing to generate 
Schr\"odinger cat states with various quantum devices, ranging from 
trapped ions and atoms to superconducting circuits. 

Viewed from a broader perspective, our results enforce the conclusions
reached in \cite{PRESKILL}, where the use of non-classical states and 
quantum resources of relevance for quantum information processing 
is also anticipated to improve our capabilities to gather information
in high precision measurements. For Schr\"odinger 
cat states in particular, the identification of contractivity features 
provides independent additional support to the idea that highly 
delocalized states may develop enhanced quantum sensitivity as recently 
suggested by Zurek \cite{ZUREK2001}. While several questions remain to 
be addressed in more depth, the present analysis thus points to a novel 
significance and potentially useful applications of Schr\"odinger cat 
states in the context of quantum-limited measurements.  

\section*{Acknowledgments} 

L.V. gratefully acknowledges current support from the Los Alamos Office of the
Director through a J. R. Oppenheimer Fellowship,
and also wishes to thank the University of Padova for hospitality at 
the time in which these ideas originated. This work was also supported 
in part by Cofinanziamento MIUR protocollo 
MM02263577${}_{-}$001.

\appendix
\section{Schr\"odinger representation}

For completeness, we report the evaluation of the position variance 
in the Schr\"odinger representation. For two-component Schr\"odinger 
cat states, the starting point is the wave function 
$$ \psi_{S2}(x,0)=N_2 \left\{k_+ \exp[-(x-x_{0})^{2}/2\Delta_0^2]+
          k_- \exp[-(x+x_{0})^{2}/2\Delta_0^2]\right\} \:, $$
where the normalization constant (also given in Eq. (\ref{normaliz})) 
is 
$$ |N_2|^{2}=( \pi \Delta_{0}^2 )^{-1/2} \Big[ |k_+|^{2}+|k_-|^{2}+
(k_+ k_-^{*}+k_-^{*}k_+) e^{-\delta^2} \Big]^{-1}\:, $$
and, as before, $\delta=x_0/\Delta_0$.  
The time evolution is obtained by solving the Schr\"odinger 
equation with effective time $\eta$, 
$$ {\partial \psi \over \partial \eta } = i {\Delta_0^2 \over 2} \,
{\partial^2 \psi \over  \partial x^2} \:, $$
which is easily accomplished by Fourier expanding the Gaussian states 
in terms of plane waves (see also \cite{MICROCAT}). The resulting 
wave function may be written as:
\begin{eqnarray*}
\psi_{S2}(x,\eta)= N_2 \, {\exp{(-{i \over 2} \text{atan}\,\eta )} 
\over (1+\eta^2)^{1/4}} 
& & \left[    
\, k_+  \exp\left(-{(1-i\eta)(x-x_0)^2 \over {2 \Delta_0(1+\eta^2)}}\right) +
\right.  \nonumber \\   
& & \left. \; \: 
\, k_- \exp\left(-{(1-i\eta)(x+x_0)^2 \over {2 \Delta_0(1+\eta^2)}}\right) 
\right]
\:.
\end{eqnarray*}
Thus, the required first and second momenta of the position observable are 
calculated as:
$$ \langle \hat{x} \rangle_{ \psi_{S2}(x,\eta) } = 
|N_2|^{2} (\pi \Delta_{0}^2)^{1/2} x_0  \left[|k_+|^{2}-|k_-|^{2}-i \eta 
(k_+ k_-^{*}-k_-^{*}k_+) e^{-\delta^2} \right] \:, 
$$
and 
\begin{eqnarray*}
\langle \hat{x}^{2} \rangle_{\psi_C(x,\eta)  } 
%\int_{-\infty}^{+\infty} |\psi(x)|^{2} x^{2} dx 
&=&
|N_2|^{2} (\pi \Delta_0^2)^{1/2} \left[ (|k_+|^{2}+|k_-|^{2})
\left( {(1+\eta^{2}) \Delta_0^2 \over 2} +x_0^2\right)
+ \right. \nonumber \\ 
&+ & 
\left. (k_+ k_-^{*}+k_+^{*}k_-)
\left( { (1+\eta^{2}) \Delta_0^2 \over 2} -\eta^2 x_0^2\right)
 e^{-\delta^2} \right] \:. 
\end{eqnarray*}
The position variance $\Delta x^2_{S2} (\eta)=
\langle \hat{x}^{2} \rangle_{\psi_{S2} (x,\eta)} -\langle \hat{x} 
\rangle^2_{\psi_{S2} (x,\eta)}$ agrees with the result 
(\ref{sigmaxt}) obtained from the Heisenberg representation.

In the case of three-component Schr\"{o}dinger cat states as considered
in Sect. IIIB, the initial wave function is 
$$ \psi_{S3} (x,0)= N_3 \left\{
k_{+} \exp[-(x-x_0)^2/2\Delta_0^2]+
k_{0} \exp[-x^2/2\Delta_0^2]+
k_{-} \exp[-(x+x_0)^2/2\Delta_0^2]
\right\} \:, $$
with $N_3$ (also quoted in Eq. (\ref{normaliz3})) given by 
\begin{eqnarray*}
|N_3|^2 & = & (\pi \Delta_0^2)^{-1/2} 
\left[ 
|k_+|^2+|k_0|^2+|k_-|^2+(k_+^* k_-+k_+ k_-^*) 
e^{-\delta^2} 
\right. \nonumber \\  & + & \left. 
(k_+^* k_0+k_+ k_0^*) e^{-\delta^2/4 }+
(k_-^* k_0+k_- k_0^*) e^{-\delta^2/4} 
\right]^{-1} \:.
\end{eqnarray*}
The time-evolved wave function $\psi_{S3}(x, \eta)$ can again be 
evaluated by solving the above Schr\"{o}dinger equation. 
After cumbersome but straightforward calculations 
we obtain the average values of the position and its square as follows:
\begin{eqnarray*}
\langle \hat{x} \rangle_{\psi_{S3} (x,\eta)} & = & |N_3|^2 
(\pi \Delta_0^2)^{1/2} x_0
\left[ c_+^2-c_-^2+ {1 \over 2} (k_+^* k_0+k_+ k_0^*)
  e^{-\delta^2/4} - {1 \over 2} (k_-^* k_0+k_- k_0^*) e^{-\delta^2/4} 
\right. \nonumber \\ & + &  \left. \left(
{i \over 2} (k_+^* k_0-k_+ k_0^*) e^{-\delta^2/4} 
+ i  (k_+^* k_--k_+ k_-^*) e^{-\delta^2} 
 -  {i \over 2} 
(k_-^* k_0-k_- k_0^*) e^{-\delta^2/4} \right) \eta \right] 
\label{xav3}
\end{eqnarray*}
and 
\begin{eqnarray*}
\langle \hat{x}^2 \rangle_{\psi_{S3} (x, \eta) } & = & 
|N_3|^2 (\pi \Delta_0^2)^{1/2}  
\left[ {1\over 2} (c_+^2+c_0^2+c_-^2) \Delta_0^2 (1+\eta^2)+
(c_+^2+c_-^2) x_0^2 
\right. \nonumber \\ & + & \left. 
{1 \over 2} (k_+^* k_0+k_+ k_0^*) e^{-\delta^2/4} \Delta_0^2 (1+\eta^2)+
{1 \over 4} (k_+^* k_0+k_+ k_0^*) e^{-\delta^2/4} x_0^2 (1-\eta^2)
\right. \nonumber \\ & + & \left. 
{i \over 2} (k_+^* k_0-k_+ k_0^*) e^{-\delta^2/4} x_0^2 \eta+
{1 \over 2} (k_+^* k_-+k_+ k_-^*) e^{-\delta^2} \Delta_0^2(1+\eta^2) +
\right. \nonumber \\ & - & \left. 
(k_+^* k_-+k_+ k_-^*) e^{-\delta^2} x_0^2 \eta^2+ 
{1 \over 2} (k_-^* k_0+k_- k_0^*) e^{-\delta^2/4} \Delta_0^2 (1+\eta^2) 
\right. \nonumber \\ & + & \left. 
{1 \over 4} (k_-^* k_0+k_- k_0^*) e^{-\delta^2/4} x_0^2 (1-\eta^2)+
{i \over 2} (k_-^* k_0-k_- k_0^*) e^{-\delta^2/4} x_0^2 \eta \right]
\:.
\label{x2av3}
\end{eqnarray*}
The position variance may be calculated from $\Delta x^2_{S3} (\eta)=
\langle \hat{x}^{2} \rangle_{\psi_{S3} (x,\eta)} -\langle \hat{x} 
\rangle^2_{\psi_{S3}(x,\eta)}$. By collecting terms of the same
order in $\eta$ and taking the ratio to the SQL value, the relative
variance $\Lambda_{S3}(\eta)$ may be cast in the form (\ref{RHOCAT3}).
The above procedure was implemented numerically to calculate the optimal
value $\Lambda_{S3}^* $ and infer the corresponding parameters. 

%%%%%%%%%%%%%%%%%%%%%%%%%%%%%%%%%%%%%%%%%%%%%%%%%%%%%%%%%%%%%%%%%%%%%%

%%%%%%%%%%%%%%%%%%%%%%%%%%%%%%%%% FIGURE CAPTIONS
\newpage

\end{document}